\documentclass[twocolumn]{aastex631}
\usepackage{makecell}
\usepackage{amsmath}

\begin{document}

\title{The Emerging Black Hole Mass Function in the High-Redshift Universe}

\author[0000-0002-6038-5016]{Junehyoung Jeon}
\affiliation{Department of Astronomy, University of Texas, Austin, TX 78712, USA}
\author[0000-0002-4966-7450]{Boyuan Liu}
\affiliation{Institut für Theoretische Astrophysik, Zentrum für Astronomie, Universität Heidelberg, D-69120 Heidelberg, Germany}
\author[0000-0003-1282-7454]{Anthony J. Taylor}
\affiliation{Department of Astronomy, University of Texas, Austin, TX 78712, USA}
\author[0000-0002-5588-9156]{Vasily Kokorev}
\affiliation{Department of Astronomy, University of Texas, Austin, TX 78712, USA}
\author[0000-0002-0302-2577]{John Chisholm}
\affiliation{Department of Astronomy, University of Texas, Austin, TX 78712, USA}
\author[0000-0002-8360-3880]{Dale D. Kocevski}
\affiliation{Department of Physics and Astronomy, Colby College, Waterville, ME 04901, USA}
\author[0000-0001-8519-1130]{Steven L.~Finkelstein}
\affiliation{Department of Astronomy, University of Texas, Austin, TX 78712, USA}
\author[0000-0003-0212-2979]{Volker Bromm}
\affiliation{Department of Astronomy, University of Texas, Austin, TX 78712, USA}
\affiliation{Weinberg Institute for Theoretical Physics, University of Texas, Austin, TX 78712, USA}
\email{junehyoungjeon@utexas.edu}

\begin{abstract}
Observations with the {\it James Webb Space Telescope (JWST)} have identified an abundant population of supermassive black holes (SMBHs) already in place during the first few hundred million years of cosmic history. Most of them appear overmassive relative to the stellar mass in their host systems, challenging models of early black hole seeding and growth. Multiple pathways exist to explain their formation, including heavy seeds formed from direct collapse/supermassive stars or sustained super-Eddington accretion onto light stellar remnant seeds. We use the semi-analytical code A-SLOTH to predict the emerging SMBH mass function under physically motivated models for both light and heavy seed formation, to be compared with upcoming ultra-deep {\it JWST} surveys. We find that both pathways can reproduce observations at $z\sim5-6$, but have distinct features at higher redshifts of $z\sim10$. Specifically, \textit{JWST} observations have the potential to constrain the fraction of efficiently accreting (super-Eddington) SMBHs, as well as the existence and prevalence of heavy seeds, in particular through ultra-deep observations of blank fields and/or gravitational lensing surveys. Such observations will provide key insights to understand the process of SMBH formation and evolution during the emergence of the first galaxies. We further emphasize the great promise of possible SMBH detections at $z\gtrsim 15$ with future {\it JWST} observations to break the degeneracy between light- and heavy-seed models.
\end{abstract}


\section{Introduction} \label{sec:intro}

It has been firmly established that most galaxies in the local Universe host a supermassive black hole (SMBH) at their centers \citep[e.g.,][]{Kormendy2013}. SMBHs are also known to accrete baryonic material from their vicinity and produce large amounts of radiation as active galactic nuclei \citep[AGN;][]{Heckman2014,Hickox2018}. The seeds of SMBHs are expected to have formed at earlier times, subsequently evolving through accretion, feedback, and merger processes to produce the local correlations seen today between the SMBHs and the stellar properties of their host galaxies \citep[e.g.,][]{Gebhardt2000,Graham2011,Beifiori2012,Croton2006,Ding2020}. To understand such coevolution, observations of SMBHs at high redshifts are crucial, and with the launch of the {\it James Webb Space Telescope (JWST)} a multitude of new AGN at $z\sim3-10$ have been discovered \citep[e.g.,][]{Kocevski2023,Kocevski2024,Ding2022,Larson2023,Matthee2023,Onoue2023,Furtak2023,Bogdan2023,Juodbalis2023,Bosman2023,Greene2023,Kokorev2023,Fujimoto2023,Maiolino2023,Taylor2024}.

These high-redshift SMBHs pose key questions: How did such massive objects form so early in cosmic history \citep[e.g.,][]{Smith2019_2,Woods2019,Inayoshi2020}? The challenge to explain the emergence of quasars with large BH masses (log\,$M_{\rm BH}/M_{\odot} >$ 8) at $z\gtrsim 6$ \citep{Wu2015,Banados2018,Zubovas2021,Fan2023} has been accentuated further with the recent \textit{JWST} observations of AGN at even higher redshifts \citep[e.g.,][]{Larson2023,Furtak2023,Greene2023}. Of particular interest is the newly-discovered, ubiquitous population of compact and highly dust-obscured objects, dubbed Little Red Dots (LRDs), that was unnoticed before \textit{JWST} \citep{Kocevski2023,Matthee2023,Kocevski2024,Akins2024}. One possible explanation for the physical nature of these objects is that their rest-optical emission is powered by dust-obscured AGN \citep{Durodola2024,Huang2024,Kokorev2024}. Other scenarios, such as extremely dense stellar clusters, have also been proposed to explain LRDs \citep{Leung2024,Guia2024,Perez2024,Baggen2024}. Surprisingly, the vast majority of newly discovered high-$z$ AGN are X-ray weak \citep[with a few exceptions e.g.][]{Kocevski2024}, with only upper limits established so far, which is in stark contrast to lower redshift AGN and quasars \citep[e.g.,][]{King2024,Pacucci2024}. Furthermore, many \textit{JWST}-discovered AGN exhibit ``overmassive" configurations, where the SMBH to galaxy stellar mass ratio is much higher than in the local Universe, subject to uncertainties in mass measurement methodology \citep[e.g.,][]{Bogdan2023,Kokorev2023,Pacucci2023,Natarajan2023}. This raises the question of whether the stellar system or the SMBH formed first, and how the two components affected each other through their respective feedback, eventually establishing the local correlations \citep[e.g.,][]{LiuBromm2022,Liu2023,Kokorev2024,Silk2024}.

Two main formation channels have been suggested to address the aforementioned observations of massive SMBHs and overmassive systems at high redshift \citep{Haemmerle2020,Inayoshi2020,Volonteri2021,Sassano2021,Regan2024}: The first invokes `light seed' remnant black holes (BHs), originating from the death of the first, metal-free, Population~III (Pop~III) stars \citep{Madau2001,Heger2003}. Lacking any efficient metal cooling in the primordial clouds, Pop~III stars are predicted to have a top-heavy initial mass function (IMF) \citep[e.g.,][]{Stacy2016,Hirano2017,Latif2022}, resulting in more massive BH remnants ($\sim10^2-10^3$ M$_\odot$) compared to local stellar remnant BHs. After formation, such Pop~III seed black holes will grow further through accretion and mergers \citep{Jeon2012,SmithRegan2018,Bhowmick2023,Trinca2022,Porras2025}.

The second, possibly less frequent, channel postulates `heavy seed' direct-collapse black holes (DCBHs), formed from the collapse of a massive extremely metal-poor \citep[$Z\lesssim 10^{-3}\ \rm Z_\odot$,][]{Chon2024} gas 
%primordial gas 
cloud, involving a supermassive star as an intermediate, short-lived stage \citep[e.g.,][]{Bromm2003,Begelman2006,Lodato2006}. This scenario, with a larger initial BH seed mass ($\sim10^4-10^6$ M$_\odot$), relies on the rare conditions that prohibit gas cooling mechanisms at low temperatures to allow the cloud to collapse without fragmenting to form multiple stars \citep[e.g.,][]{Lodato2007,Johnson2013,Wise2019,Haemmerl2018,Haemmerle2020,Luo2020}. Alternatively, even if fragmentation is not completely avoided, heavy BH seeds can also form through runaway collisions of (proto-)stars or BHs in dense stellar clusters with high gas inflow rates and rapid accretion flows \citep{Zwick2023,Reinoso2023,Klessen2023,Gaete2024}. Starting with a larger initial mass, such DCBH seeds may be subject to weaker timing constraints in growing to the observed high-$z$ SMBH masses \citep[e.g.,][]{Haiman2001}. Furthermore, the heavy seed channel could naturally explain the inferred overmassive systems by forming a massive SMBH in an environment with initially few stars \citep{Durodola2024,Jeon2024}.

After BH seed formation, there are different scenarios for their growth to become the observed massive AGN. The theoretical maximum AGN luminosity is the Eddington limit, with a corresponding fiducial accretion rate that depends on the physics of the accretion flow \citep{Begelman1979}. The light seed channel under Eddington-limited growth cannot produce the observations of some of the most massive high redshift AGN observations, due to ineffective growth over extended periods from stellar feedback that is heating the gas and wandering movements of the central BH \citep{Milosavljevic2009,Johnson2007_2,Jeon2022,Partmann2024}. Heavy seed channels, on the other hand, can reach the inferred SMBH masses, as they operate in environments with less-prominent stellar components, with more stationary SMBHs due to their larger initial masses \citep{LiHernquist2007,Inayoshi2020,Larson2023,Jeon2024}. However, as the Eddington limit assumes spherical accretion, super-Eddington accretion rates are possible, at least for extended periods, via geometrically thick disk accretion modes \citep{Jiang2014,Jiang2019,Davis2020,Safarzadeh2020}. Invoking such super-Eddington accretion, the light seed channel could also give rise to the high-redshift SMBHs, possibly providing a natural explanation for the unusual X-ray weakness of the AGN observed by \textit{JWST} \citep[e.g.][]{Maiolino2024}, including the LRDs \citep{Pacucci2024,King2024,Madau2024,Inayoshi2024,Inayoshi2025}. 

To distinguish between the possible pathways towards the first SMBHs, further observations are needed that reach higher redshifts, supported by targeted theoretical predictions to interpret the observations. The heavy-seed model in particular can be independently tested with gravitational-wave (GW) detections of merging SMBH binaries with the Laser Interferometer Space Antenna (LISA; \citealt{Robson2019}), capable of accessing the required low-frequency regime \citep[e.g.,][]{Liu2020}. Moreover, pulsar timing arrays (PTAs) have found evidence for a stochastic gravitational wave background (GWB) \citep{Agazie2023,Antoniadis2023,Reardon2023,Xu2023}, that could be partially sourced by binary SMBHs \citep{Hobbs2017,Romano2017}. However, the PTA constraints on the SMBH population are currently weak due to the high noise of the initial GWB data \citep[e.g.,][]{Agazie2023_2,Agazie2023_3}. 

In principle, the BH mass function (BHMF) at high-$z$ could be a powerful probe to disentangle distinct seeding models. \citet{Taylor2024} recently measured the broad-line AGN (BLAGN) BHMF over a wide dynamic range of BH masses for the first time at $z\sim5$. However, comparing to models with a variety of seeding and growth mechanisms, they found that multiple models were consistent with the observations. They concluded that by $z\sim5$, the ``memory" of BH seeding is lost, and that BHMF measurements at higher redshifts (closer to the BH origin epoch) are needed. 

Therefore, in this work, we aim to study the population statistics of even higher-redshift SMBHs, providing predictions for the evolving mass function and their hosts at $z\gtrsim 9$ under different SMBH evolution scenarios. We specifically utilize the semi-analytic code (SAM) A-SLOTH (Ancient Stars and Local Observables by Tracing Halos) \citep{Hartwig2022,Magg2022,Hartwig2024}, that models the formation and evolution of the first stars and is tuned to high-redshift constraints. We develop SMBH/AGN formation and growth models to be used in the A-SLOTH framework to study the co-evolution of the SMBH, stellar, and halo populations. A-SLOTH is highly efficient and parallelized so that different SMBH scenarios can be rapidly tested and compared. Unlike previous works, we use a SAM focused on the high-redshift regime that models the formation of the first stars so that the first BH formation and evolution in the early Universe can be more accurately followed. 

This paper is organized as follows. In Section \ref{sec:methods}, we introduce A-SLOTH, its relevant features, and the SMBH models we have developed. 
%We compare the results against existing observations and validate our models in Section \ref{sec:calibration}. 
We present our predictions for different SMBH evolution scenarios and their differences in Section \ref{sec:results}. In Section \ref{sec:discussion} we discuss our results in the context of overall SMBH evolution at high redshifts and assess the prospects for future SMBH observations. We summarize our findings in Section \ref{sec:conclusions}.

% nanograv? 

\section{Methodology} \label{sec:methods}

\subsection{A-SLOTH} \label{sec:asloth}
A-SLOTH is a semi-analytical framework to model high-redshift galaxy formation, based on halo merger trees from N-body simulations or the Extended Press-Schechter (EPS) formalism \citep{Parkinson2008}. The code has been calibrated to well reproduce the cosmic star formation rate density at $z\sim 4.5-13.3$ \citep{Hartwig2024}. We here employ the EPS approach, utilizing its computational speed to explore a broad parameter space. Within the provided merger tree, A-SLOTH models the formation and evolution of stellar populations and their effects on the cosmic environment. A-SLOTH uses an adaptive timestep to trace halo evolution which may be smaller than the time between two levels in the underlying tree. This timestep is set to be a small fraction of the minimum between the halo star formation, accretion, dynamical, and merger tree timescales \citep[see sec.~2.4 in][]{Hartwig2022}, with a typical value around $0.5\sim0.01$ Myr. Below, we briefly summarize the relevant star formation model in A-SLOTH, as well as our code modifications made to model SMBH evolution, and refer the reader for full details to the public release papers \citep{Hartwig2022,Magg2022}, as well as to the previous implementations of nuclear star clusters (NSCs) and galactic dynamics in \citet{Liu2024}.

\subsubsection{Star formation}\label{sec:sf}
A-SLOTH sets a critical halo mass above which the gas in the halo is assumed to be able to rapidly cool, such that star formation (SF) can occur. This critical mass is the minimum between the atomic cooling threshold where the halo acquires a virial temperature of $T_{\rm vir}\geq10^4$\,K, and the molecular cooling threshold that depends on the large-scale streaming velocity of the baryons ($v_{\rm BC}$) as well as the global Lyman-Werner (LW) background defined as 
\begin{equation}\label{globallw}
    J_{21,\rm global} = 10^{2-z/5} \mbox{\ ,}
\end{equation}
where the background is given in units of $J_{21}$ or $10^{-21}$ erg s$^{-1}$ cm$^{-2}$ Hz$^{-1}$ sr$^{-1}$ \citep{Greif2006,Hartwig2022}. The critical halo mass for efficient molecular cooling is \citep{Schauer2021}
\begin{equation}
    \log_{10}(M_{\rm crit}/M_\odot) = 6.02(1.0+0.17\sqrt{J_{21}})+0.42v_{\rm BC}\mbox{.} 
\end{equation}
If the halo mass exceeds either of the critical mass values, it is considered for SF. \citet{Hartwig2022} chose $v_{\rm BC} = 0.8$ as the most probable value \citep{Schauer2019}, for which the two masses are equal at $z\lesssim10$, also used here as our default parameter choice. The $v_{\rm BC}$ value is normalized by the rms streaming velocity at recombination, $\sigma_{\rm rms}\simeq30$ km s$^{-1}$ \citep{Schauer2021}. However, in reality, there is a distribution of the streaming velocity $v_{\rm BC}$ encountered by the dark matter halos. Therefore, we also test two other $v_{\rm BC}$ values, 0 and 2, and create a secondary model from the weighted mean of the three runs with these three $v_{\rm BC}$ values. The weighted mean is determined from the probability distribution of $v_{\rm BC}$ \citep{Liu2024_2}. 

When a halo forms stars, if the halo metallicity is below a critical value, the first generation of metal-free stars (Pop III) will be formed and if the metallicity is above the critical value, the second generation of metal-enriched stars (Pop II) will be formed. This critical metallicity is defined as  \begin{equation}
    10^{\rm  [C/H]-2.30}+10^{\rm [Fe/H]}>10^{-5.07}
\end{equation}
where [Fe/H] and [C/H] are the iron and carbon abundances of the star-forming gas such that when this condition is met, Pop II stars will form \citep{Chiaki2017}. Furthermore, A-SLOTH divides the baryonic material in the halo into four components: cold gas, hot gas, stars, and outflows, where only the cold gas contributes to SF. The newly formed stellar population is assigned a mass based on the cold gas properties according to
\begin{equation}
    M^i_{*} = \eta_*M^i_{\rm cold}\frac{\delta t_i}{t^i_{\rm cold,ff}}\mbox{\ ,}
\end{equation}
where $i$ is the timestep index, $M_{\rm cold}$ the cold gas mass of the halo, $t_{\rm cold,ff}$ the corresponding free-fall time, $\delta t$ the timestep, and $\eta_*$ the star formation efficiency (SFE). The SFE has been calibrated in \citet{Hartwig2024} to have best-fit values of 8.15 for Pop III and 0.237 for Pop II stars. The SFE can be greater than 1 as it is defined as the fraction of cold gas that is converted into stars per free fall time for the average cold gas density. An SFE larger than 1 thus represents the case when the star formation timescale is shorter than the average free-fall time. The average number of stars is computed in logarithmically spaced IMF bins, which is used to draw individual stars with Poisson sampling. For Pop II, the Kroupa IMF \citep{Kroupa2001} is used and for Pop III, a power law IMF of
\begin{equation}
    \frac{dN}{d\log M_{\rm star}} \propto M^{-1}_{\rm star}
\end{equation}
is used. The Pop~III component has a stellar lifetime based on the fitting function \citep{Schaerer2002}
\begin{equation}
    \log_{10}(t_{\rm III}/ {\rm yr}) = 9.785 - 3.759x+1.413x^2-0.186x^2\mbox{\ ,}
\end{equation}
whereas for Pop~II, the lifetime is expressed as \citep{Stahler2004}
\begin{equation}
    \log_{10}(t_{\rm II}/ {\rm yr}) = 10-3.68x+1.17x^2-0.12x^3\mbox{\ ,}
\end{equation}
where $x = \log_{10}(M_{\rm star}/M_\odot)$ and $M_{\rm star}$ is the individual stellar mass.

\subsubsection{Photoheating Feedback} \label{sec:photofeedback}

Photoheating feedback from massive stars ($>5$ M$_\odot$) can convert the cold gas in halos to hot gas. The instantaneous conversion rate is estimated assuming the \citet{Spitzer1978} solution for the HII region expansion from stellar photoionization as 
\begin{equation}
    \dot{M}_{\rm heat} = m_{\rm H}n_{\rm cold}R^2_{\rm St}c_{s, \rm ion}\left[1+\frac{7}{4}\frac{c_{s, \rm ion}(t-t_{\rm St})}{R_{\rm St}}\right]^{-1/7}
\end{equation}
where $m_{\rm H}$ is the hydrogen mass, $n_{\rm cold}=10^3$ cm$^{-3}$ the neutral gas number density around a newly formed star, $R_{\rm St}$ the Str$\ddot{\rm o}$mgren radius, $t$ the time, $t_{\rm St}$ the time the ionization front reaches $R_{\rm St}$, and $c_{s, \rm ion}$ the sound speed of the ionized gas at $10^4$ K. 

A-SLOTH assumes that 90\% of massive stars in a halo form inside one cluster at the galactic center, whereas the other 10\% of massive stars form in isolation \citep{Chen2022}. As regions of heating overlap for star clusters, they heat up their environment less efficiently compared to isolated stars. Therefore, the ionizing photons from 90\% of massive stars are combined to compute one mass conversion rate for the star cluster and individual rates are computed for isolated stars. In a time step $dt$, the total gas mass heating rate from massive stars is thus given by
\begin{equation}
    \delta M_{{\rm heat}} = \left(\dot{M}_{\rm heat, cluster}+\sum^{0.1N}_{j=1}\langle \dot{M}_{\rm heat, isolated}^j \rangle \right)dt\mbox{,}
\end{equation}
where $N$ is the number of massive stars.
 
\subsubsection{Supernova Feedback} \label{sec:snefeed}

After a stellar lifetime has passed, if the star's mass is in the range $10-40$ M$_\odot$, it will explode as a core-collapse supernova, and if a Pop~III star is in the mass range $140-260$ M$_\odot$, it will explode as a pair-instability supernova. The energies produced by these events are shown in figure~2 of \citet{Hartwig2022}. The feedback from supernovae (SNe) is implemented as gas ejections from the halo by comparing the gas binding energy to the SN feedback energy, where the former has contributions from dark matter (DM), cold and hot gas, as well as stars. Specifically, the hot and cold gas binding energies at step $i$ are defined as 
\begin{multline}
    E^i_{\rm bind,hot} = \frac{GM_{\rm vir, peak}M_{{\rm hot},i}}{R_{\rm vir}}\chi_{\rm hot, 1} + \chi_{\rm hot, 2}
\end{multline}

\begin{multline}
    E^i_{\rm bind,cold} = \frac{GM_{\rm vir, peak}M_{{\rm cold},i}}{R_{\rm vir}}\chi_{\rm cold, 1} + \chi_{\rm cold, 2}
\end{multline}
where $M_{\rm vir,peak}$ is the peak halo virial mass, $R_{\rm vir}$ the halo virial radius, $M_{\rm hot, cold}$ the hot/cold gas mass, and $\chi_{\rm (hot,cold), (1,2)}$ are factors accounting for the binding energy arising from the dark matter, the hot/cold gas disk, and the stellar disk \citep{Chen2022}. Full expressions and derivations for these binding energies can be found in \citet{Chen2022,Hartwig2022}.

The total SN energy at each step $i$, $E^i_{\rm SNe}$, is distributed to some fractions of hot and cold gas, depending on the respective binding energies as 

\begin{align}
    f_{\rm hot} = \frac{E^i_{\rm bind,hot}M^i_{\rm hot}}{E^i_{\rm bind,hot}M^i_{\rm hot}+E^i_{\rm bind,cold}M^i_{\rm cold}} \\
    f_{\rm cold} = \frac{E^i_{\rm bind,cold}M^i_{\rm cold}}{E^i_{\rm bind,hot}M^i_{\rm hot}+E^i_{\rm bind,cold}M^i_{\rm cold}} \mbox{\ .}
\end{align}
Furthermore, we adopt an outflow efficiency $\gamma_{\rm out}$, defined as \citep{Chen2022}
\begin{equation}
    \gamma_{\rm out} = \left(\frac{M_{\rm vir,peak}}{M_{\rm out,norm}}\right)^{\alpha_{\rm out}}\mbox{\ .}
\end{equation}
The normalization mass, $M_{\rm out,norm}$, and $\alpha_{\rm out}$ are free parameters calibrated in \citet{Hartwig2024} so that for the same SN energy, relatively more gas is removed from less massive halos than more massive ones.

Overall, the amount of hot/cold gas ejected by SNe is determined as the ratio between the SN energy and the gas binding energy, accounting for the outflow efficiency and the fraction of gas to be affected, according to

\begin{multline}\label{sneoutflow}
    \delta M^i_{\rm out,(hot,cold)} = \text{min} \biggl(\frac{E^i_{\rm SNe}f_{(\rm hot,cold)}/\gamma_{\rm out}}{E_{\rm bind,(hot,cold)}}M^i_{\rm (hot,cold)},
    \\M^i_{\rm (hot,cold)}\biggr)
\end{multline}
The SN metal yields are taken from the tabulation in \citet{Nomoto2013} for Pop~III and \citet{Kobayashi2006} for Pop~II stars.

\subsection{Black hole seeding} \label{seeding}

We consider two classes of BH seeds, light ones originating from Pop~III stellar remnants ($\sim100$\,M$_\odot$) and heavy DCBH seeds from collapsing massive clouds/supermassive stars ($\sim10^5$\,M$_\odot$). Our specific process of assigning BH seeds to halos is as follows:

Each halo in the merger tree is checked to see if any of the halo's progenitors hosted a BH already. If no progenitors hosted a BH/if the halo is the very first progenitor, we assess the halo conditions and determine whether it should be seeded with a BH. When a massive ($>40\ \rm M_\odot$) Pop III star in the halo dies and its mass is outside the core-collapse or pair-instability SN range, $40$ M$_\odot<M_*<140$ M$_\odot$ or $M_*>260$ M$_\odot$, a light seed of the same mass as the dying Pop III star is assigned at the halo center. We do not consider Pop II stars for simplicity and as they will result in much lower mass BHs ($\sim5-10$ M$_\odot$) compared to the more massive Pop~III remnants \citep{Stacy2016,Volonteri2021,Latif2022,Sassano2021}. 

For heavy seeds, we consider a set of criteria based on the halo virial temperature, metallicity, and LW feedback to capture the dense, hot, and metal-poor conditions required for DCBH formation and ensure that the gas in the halo is not able to cool and fragment too quickly to form regular stars instead \citep{Bromm2003,Ardaneh2018,Wise2019,Chon2021}. Specifically, we require that the virial temperature of the halo be greater than $10^4$\,K, above the atomic cooling limit, to be able to host gas that can collapse (nearly) isothermally, even in the absence of H$_2$ cooling. We further impose that the metallicity of the star-forming gas in the halo be smaller than a critical metallicity, $Z<Z_{\rm crit} = 2\times10^{-4}~{\rm Z}_\odot$ \citep{Liu2020} to not allow for too-efficient metal cooling. Recent work has shown that heavy seeds could form even at higher metallicities up to $10^{-3}~{\rm Z}_\odot$ \citep{Chon2024}. We have tested this higher metallicity threshold, and found that while around 50\% more heavy seeds do form, they are still subdominant to the overall BH population and do not significantly affect our results especially at higher redshifts. Thus, we adopt the default critical value of $2\times10^{-4}~{\rm Z}_\odot$ throughout the paper. Finally, we impose that the LW background in the halo be greater than the critical level, $J_{\rm LW}>J_{\rm crit}$, so that LW radiation can dissociate H$_2$ and disable molecular cooling, thus preventing star formation. 

We set $J_{\rm crit}=300$, in units of $J_{21}$ \citep{Trinca2022}, and we consider both global and local LW contributions\footnote{The physics behind the critical flux is complex and still rather uncertain \citep[see, e.g.,][]{Sugimura2014}. The resulting DCBH number density is thus very uncertain as well, varying with the critical level as $J_{\rm crit}^{-4}$ \citep[e.g.,][]{Inayoshi2015, Chon2016}.}. The global LW background is defined in Equation~\ref{globallw} above, and is generally subdominant relative to $J_{\rm crit}$, but we include it for completeness. The local LW flux within a given halo generally provides the main contribution to the LW radiation, and is calculated from considering massive stars, above 5\,M$_\odot$, that are capable of producing LW radiation ($11.2-13.6$~eV) efficiently. We determine the LW photon production rate of each active massive stars based on the stellar mass from the fitting formula in \citet[][see their equ.~8]{Deng2024}, further assuming for simplicity that the high-mass stars are located on average at 0.1$R_{\rm vir}$ of the halo center. Thus, for a given halo, its LW flux is the sum of the global background and the local component, produced by the massive stars in the halo. 
Other studies have concluded that local LW radiation alone could not establish conditions for DCBH formation \citep{Sullivan2025}, as gas must initially cool below $\lesssim 1,000$\,K to form stars first, and there is insufficient time to subsequently heat the gas up again to the atomic-cooling threshold. Our model represents the optimistic scenario where the local massive stars have formed in the progenitor halos, whose mergers heat the gas to $\sim 10^4$~K, triggering prompt DCBH formation within $\sim 1$~Myr, before the gas reservoir is destroyed by feedback. If all the above criteria are met, regarding virial temperature, metallicity, and LW radiation, the halo is endowed at its center with a heavy seed of mass $10^5$ M$_\odot$ \citep[e.g.,][]{Becerra2018b,Becerra2018a}. The median cold gas mass in halos right after DCBH formation is $\sim5\times10^{4}$ M$_\odot$, of the same order as the initial DCBH mass, agreeing with the theoretical scenario that DCBH formation should take up most of the initially available cold gas in the host halo \citep{Wise2019}.

If any of the halo's progenitors contains a BH, the halo inherits at its center the one from the most massive progenitor host. If the most massive progenitor hosts multiple BHs, the other BHs are also inherited at their respective positions. If multiple progenitors host BHs, the BHs from the less-massive progenitors are inherited as well, but placed at random (apocenter) distances from the halo center. In assigning distances, we follow the spatial distribution of Pop~III remnants, derived from high-resolution simulations \citep{Liu2020,Liu2020_2} for BHs with masses $M_{\rm BH} < 10^5$\,M$_{\odot}$, and the locations of nuclear star clusters (NSCs) after halo mergers found in previous A-SLOTH implementations \citep{Liu2024} for BHs with $M_{\rm BH} \geq 10^5$\,M$_{\odot}$. We have adopted the NSCs as tracers of post-merger massive BH locations, as NSCs are thought to reside in the centers of dwarf halos, similar to the massive BHs \citep{Partmann2024,Askar2024,Chen2024}. Finally, the BH orbits are assigned random eccentricities drawn from a uniform distribution in $[0,1)$ \citep{Liu2024}.

\subsection{Two models of black hole accretion} \label{sec:accretion}

At each timestep, we update the BH mass and location through accretion and dynamical friction. These steps are crucial to be able to model BH evolution, but with the lack of information on the gas distribution near the BH in SAMs, it is difficult to estimate BH accretion. Therefore, we consider two models of BH accretion, Eddington and Bondi. 

We constrain the accretion rate to be limited by the available cold gas mass in the halo as
\begin{equation}\label{eq:macc}
    \delta M_{\rm BH} = \text{min}\left(f_{\rm duty}\dot{M}_{\rm acc}dt, M_{\rm cold}\right)\mbox{\ ,}
\end{equation}
where $f_{\rm duty}$ is the duty cycle for active accretion onto the SMBH \citep{Pacucci2023,Lai2024}. Here, the duty cycle is a free parameter which we choose to reproduce existing observations (see Table~\ref{models}). For the two models, we use different methods to determine the accretion rate $\dot{M}_{\rm acc}$.

In the Eddington mode, we use the fiducial physical upper limit of accretion, the Eddington rate, thus representing optimistic BH growth trajectories. More specifically, this rate is parameterized by the Eddington ratio $f_{\rm Edd}$ as
\begin{equation}
\dot{M}_{\rm Edd} = 2.7\times10^{-3}\left(\frac{M_{\rm BH}}{10^5~\text{M}_\odot}\right)\left(\frac{\epsilon_r}{0.1}\right)^{-1}\rm~M_\odot~\text{yr}^{-1}\mbox{\ ,}
    \label{edd}
\end{equation}
such that
\begin{equation}
    \dot M_{\rm acc} =  f_{\rm Edd}\dot{M}_{\rm Edd}
        \label{edd_acc}\mbox{\ .}
\end{equation}
Here, $\epsilon_r=0.1$ is the radiative efficiency, and we allow the Eddington ratio $f_{\rm Edd}$ to be larger than 1, corresponding to super-Eddington accretion. We adjust this parameter to ensure that the resulting SMBH population agrees with the observed high-redshift BH mass function (see Section \ref{sec:results}). The Eddington model, while positing that every BH will accrete at the same fraction, $f_{\rm Edd}$, of the Eddington rate, does not make assumptions about the gaseous environment near the BH, and is mainly dependent on the current BH mass. 

We note that the radiative efficiency may be smaller (0.01-0.05) at earlier times, especially for super-Eddington cases \citep{Jiang2014,Jiang2019}. If so, the Eddington accretion rate of the SMBHs will be higher for a given mass, implying faster growth. However, the effect of lower $\epsilon_r$ can be similarly reproduced by adjusting $f_{\rm Edd}$ to a higher value, to a more super-Eddington accretion. Thus, the choice of $\epsilon_r$ is not crucial within our modeling, as the consequence of super-Eddington accretion can be represented by other free parameters.

For the Bondi model, we use the Bondi-Hoyle formalism \citep{Bondi1944}:
\begin{equation}\label{bondi}
\dot{M}_{\rm Bondi} = \alpha\frac{4\pi(GM_{\rm BH})^2\rho_g}{c_s^3}\mbox{\ ,}
\end{equation}
\\
where $\rho_g$ is the gas density, $c_s$ the sound speed, and $\alpha$ the boost factor, which is a free parameter. The boost factor accounts for the enhanced gas density in the inner regions of the halo near the central BH that is not well captured in cosmological simulations \citep{Jeon2022}, or with idealized halo profile models \citep{Trinca2022}. We set it to unity in this work, given that A-SLOTH explicitly models the cold gas in the halo. For $\rho_g$, we consider only the cold gas mass $M_{\rm cold}$ as contributing to BH accretion, assuming that it is confined to within the halo scale radius, $R_s = R_{\rm vir}/c_{\rm DM}$, of the Navarro-Frenk-White (NFW) dark matter halo profile \citep{Navarro1996}, where $c_{\rm DM}$ is the halo concentration parameter \citep{Hartwig2022}. The DM concentration is given by the fitting functions from \citet{Correa2015}. Thus, we approximate the cold gas density distribution as an isothermal sphere with a flat core \citep{Trinca2022} 
\begin{equation}\label{isothermal}
    \rho(r) = \frac{\rho_{\rm norm}}{1+(r/r_{\rm core})^2}\mbox{\ ,}
\end{equation}
where $r_{\rm core} = 0.012R_{\rm vir}$ is the halo core radius \citep{Trinca2022} and $\rho_{\rm norm}$ the normalization density. The latter is set so that the integral of Equation~\ref{isothermal} up to the scale radius $R_s$ equals the total $M_{\rm cold}$. 

We evaluate this expression at the Bondi radius of the BH, $r_b = GM_{\rm BH}/c_s^2$, where $c_s = \sqrt{k_{\rm B}T/m_g}$, with $k_{\rm B}$ being the Boltzmann constant, $T$ the gas temperature, and $m_g$ the mean molecular weight of the gas. To estimate the gas temperature in an idealized fashion, we use the halo viral temperature plus an effective contribution expressing the heating from BH feedback, $T_{\rm vir} + T_{\rm feed}$, when the average metallicity is below $Z_{\rm crit}$. If the latter is above $Z_{\rm crit}$, on the other hand, we employ the cold gas temperature at the given redshift plus the BH feedback contribution, $T_{\rm cold} + T_{\rm feed}$, reflecting the fact that at high redshifts, the cold gas in the halo should be able to efficiently cool to temperatures lower than the halo virial temperature. The cold gas temperature $T_{\rm cold}$ is set to the cosmic microwave background (CMB) temperature \citep[e.g.,][]{Safranek2016}. Such cold gas near the CMB temperature has been found in high-resolution simulations even at lower redshifts $z\sim5-6$ \citep{Liu2020,Jeon2023}. We note that thus using the CMB temperature sets the upper bound on accretion.

The additional heating from BH feedback, expressed in the equivalent $T_{\rm feed}$, is described below (see Section~\ref{sec:feedback}). We further limit the accretion rate to a multiple of the Eddington value (Equ.~\ref{edd}) as 
\begin{equation}
    \dot M_{\rm acc} = \text{min}\left(\dot{M}_{\rm Bondi}, f_{\rm Edd}\dot{M}_{\rm Edd}\right)\mbox {\ .}
    \label{bondi_acc}
\end{equation}
Compared to the Eddington model, the Bondi model has the advantage of adapting to the physical conditions in the vicinity of individual BHs. However, unlike for the Eddington model, idealized estimates have to be used for the cold gas density and temperature, as such information is not directly available in SAMs. Overall, the Eddington model proceeds with fewer assumptions, whereas the Bondi model represents a more physically realistic approach.

For both models, $\delta M_{\rm BH}$ is removed from the cold gas reservoir of the halo. Accretion is only applied to the primary BH, which is assumed to reside in the halo center, where the dense and cold gas is also located. The other BHs are assumed to not accrete for simplicity, in line with previous work showing that wandering BHs that do not reside in the dense central region generally do not accrete efficiently and thus remain dormant \citep{Jeon2023,Ogata2024}. 

\subsection{Black hole dynamics and mergers} \label{sec:dynamics}

If a BH is not at the center of a halo after halo mergers, we follow its inspiral and update its location and eccentricity at each (star formation) timestep using the dynamical friction prescription from stars and DM in \citet[][their equations 2-4]{Liu2024}. Specifically, the change in distance from the center $r$ and eccentricity $e$ is modeled according to 

\begin{equation}
    \frac{dr}{dt} \simeq -r \left[\frac{1}{\tau_{{\rm DF}, *}}+\frac{1}{\tau_{{\rm DF}, \chi}}\right]
\end{equation}
\begin{equation}    
        \frac{de}{dt} \simeq -e \left[\frac{1}{\tau_{{\rm DF}, *}}+\frac{1}{\tau_{{\rm DF}, \chi}}\right] \mbox{\ ,}
\end{equation}
where $\tau_{\rm DF, *}$ is the dynamical friction timescale arising from the halo stellar component, and $\tau_{\rm DF, \chi}$ that from the DM component. This timescale can be evaluated using the Chandrasekhar formula \citep{Binney2008}
\begin{equation}\label{dftime}
    \frac{\tau_{\rm DF}}{\rm Myr} = \frac{342}{\ln \Lambda}\left(\frac{r}{3 ~\rm pc}\right)^2\left(\frac{v}{10~ \rm km~s^{-1}}\right)\left(\frac{M_{\rm BH}}{100~\rm{M}_\odot}\right)^{-1}
\end{equation}
where $\ln \Lambda$ is the Coulomb logarithm and $v$ is the circular velocity. We use $\ln \Lambda\sim\ln[M_{*}r/(0.8M_{\rm BH}R_{*})]$ and $v\sim\sigma_*\sim\sqrt{GM_*/(0.8R_*)}$ for $\tau_{\rm DF,*}$, given the total stellar mass $M_*$ and size $R_*$ of the galaxy. The stellar parameters, $M_*$ and $R_*$, are replaced with the virial mass and radius, $M_{\rm vir}$ and $R_{\rm vir}$, when calculating $\tau_{\rm DF,\chi}$. When the inner slope of the density profile of the host galaxy is $0<\gamma_*<2$, as defined in \citet{Arcasedda2015} for a dwarf starburst galaxy, we use a generalization of the Chandrasekhar formula above for evaluating the stellar term. The analytical fit used is valid for both cored and cusped density profiles and generally produces smaller timescales than Equation~\ref{dftime} \citep{Liu2024}. For a more complete description, we refer the reader to \citet{Arcasedda2015,Arcasedda2016,Liu2024}. 

We introduce a merging radius, $r_{\rm merge}$, such that if a BH wanders inside the $r_{\rm merge}$ of the central BH, we assume that the two BHs have merged, updating the mass of the central BH accordingly and removing the merged BH from subsequent tracking. The merging radius is set to 
\begin{equation}
    r_{\rm merge} = \frac{GM_{\rm BH}}{v_{\rm vir}^2}\mbox{\ ,}
\end{equation}
where $v_{\rm vir}$ is the halo virial velocity. We note that here we ignore the delay time of BH mergers proceeding under gravitational wave emission, and the gravitational recoil after merger. Therefore, we explore the optimistic case where BH mergers occur efficiently and do not remove the product from the host galaxy. 

\begin{deluxetable*}{ccccc}[htb!]
\tabletypesize{\footnotesize}
\tablenum{1}
\tablecolumns{1}
\tablecaption{Summary of model parameters  \label{models}}
\tablewidth{0pt}
\tablehead{
\colhead{Name} & \colhead{BH Seeding}& \colhead{Accretion Mode} & \colhead{$f_{\rm Edd}$} &\colhead{$f_{\rm duty}$} 
}
\startdata
Light seeds forced super-Edd & Light only & Eddington&1.5 & 0.8\\
Light seeds super-Edd-limited &Light only& Bondi &1.5 & 0.5\\
Heavy seeds Edd-limited &Heavy and Light& Bondi &1 & 0.5\\
Heavy seeds super-Edd-limited &Heavy and Light& Bondi &1.5 & 0.5\\
Heavy seeds forced super-Edd &Heavy and Light& Eddington &1.5 & 0.5\\
\enddata
\tablecomments{The Heavy seeds forced super-Edd model is only used to probe the most extreme BH mass growths (see Fig.~\ref{fig:bhmost}), as it exceeds observations even more extremely than the Light seeds forced super-Edd model (see Fig.~\ref{fig:bhmf}).}
\centering
%\endcentering
\end{deluxetable*}

\subsection{Black hole feedback} \label{sec:feedback}
%\textbf{Not implemented}

%For BH feedback, we follow a similar prescription as SNe feedback as described in section \ref{sec:snefeed}. We use equation \ref{sneoutflow} to determine the gas outflow caused by BH feedback except that the SNe energy, $E_{\rm SNe}$, is replaced with the BH radiative energy
%\begin{equation}
%    E_{\rm BH} = \epsilon_w\epsilon_r\dot{M}_{\rm BH}c^2
%\end{equation}
%where $\dot{M}_{\rm BH}$ is the total BH accretion rate, $c$ the speed of light, $\epsilon_r=0.1$ the radiative efficiency, and $\epsilon_w=0.2$ the gas coupling efficiency \citep{Nelson2019}.

%\textbf{Implemented}

We implement BH feedback as injection of thermal energy, resulting from BH accretion. More specifically, the energy injected to the nearby cold gas at each timestep $dt$ is 
\begin{equation}
     E_{\rm BH} = \epsilon_w\epsilon_r\dot{M}_{\rm BH}c^2dt\mbox{\ ,}
\end{equation}
where $\dot{M}_{\rm BH}$ is the effective BH accretion rate, and $\epsilon_w=0.02$ the radiation-thermal coupling efficiency \citep{Tremmel2017}. This injected energy is converted to the increase in the gas temperature nearby the BH according to 
\begin{equation}
    T_{\rm feed} = (\gamma-1) \frac{E_{\rm BH}m_g}{k_{\rm B} M_{\rm gas}}\mbox{\ ,}
\end{equation}
where $\gamma=5/3$ is the polytropic index of the gas, and $M_{\rm gas}$ the combined mass of the cold and hot gas in the halo. This temperature is used in determining the sound speed for the Bondi model in Equation~\ref{bondi} (see Section~\ref{sec:accretion} above). We assume that the BH feedback does not significantly affect star formation, as its impact is still quite uncertain. Some observations, for example, show no correlation between the star formation rate and AGN activity \citep[e.g.,][]{Suresh2024,Scholtz2024,Oio2024}. Furthermore, the relatively lower-mass BHs expected at high redshifts imply an overall smaller effect on star formation.

\subsection{Cosmological Volume}

To model a population of SMBHs in a cosmologically representative volume in EPS trees, we use the methodology in \citet{Magg2016}. We run A-SLOTH for 300 halo masses at $z=1$ between $M_{\rm halo, min} = 5\times10^8~M_\odot$ and $M_{\rm halo, max} = 2\times10^{13}~M_\odot$ at evenly spaced logarithmic mass bins. This range of halo masses was chosen so that the low mass end of the merger tree will still contain Pop III star forming halos and the upper mass end to include halos that are too rare to contribute significantly to a cosmologically representative sample, thus probing both the general population and extreme cases of SMBH formation and evolution. For each halo mass $M_i$, we define $M_{i, {\rm up/down}} = 0.5(M_i+M_{i\pm1})$. We then include the results of each run depending on the comoving halo number density of that mass at $z=1$ with the weight 
\begin{equation}
    w_i = \frac{f_{\rm duty}}{M_i}\int^{M_{i, {\rm up}}}_{M_{i, {\rm down}}} \frac{dn}{d\ln M}dM\mbox{\ ,}
\end{equation}
where $dn/d\ln M$ is the halo mass function at $z=1$, using the \citet{Sheth2001} halo mass function implemented in the Colossus package \citep{Diemer2018}. The factor of $f_{\rm duty}$, the fraction of the time that SMBHs are accreting, is included to account for the fact that observations will only detect actively accreting SMBHs. 
 
%\section{Model Calibration} \label{sec:calibration}

\section{Results} \label{sec:results}

We carry out a suite of runs with different sets of seeding and accretion models. For seeding, we consider two cases with heavy seeds included, and one where only light seeds form. For accretion, we explore the Eddington and Bondi models (see Section~\ref{sec:accretion}). As free parameters, we vary the Eddington ratio $f_{\rm Edd}$ and the accretion duty cycle $f_{\rm duty}$. We note that within the Eddington model, $f_{\rm Edd}$ determines the actual accretion rate for all BHs, while for the Bondi model, $f_{\rm Edd}$ sets the upper limit for accretion. We assign values to the free parameters so that the observed BH mass function at $z\sim3.5-6$ can be reproduced \citep{Taylor2024}, or the highest-redshift AGNs, at $z\sim10$, can be explained \citep{Bogdan2023,Maiolino2023}. The full suit of models explored in this paper are summarized in Table~\ref{models}. Regarding the effect of relative DM-baryon streaming, we specifically consider two sets of runs for each model, as described in Section~\ref{sec:sf}: one with $v_{\rm BC}=0.8$ for all halos, and the other with a weighted distribution of $v_{\rm BC}$ values.

\subsection{SMBH demographics}\label{sec:demographics}

\begin{figure*}[!htb]
    \centering
    \includegraphics[width=0.9\textwidth]{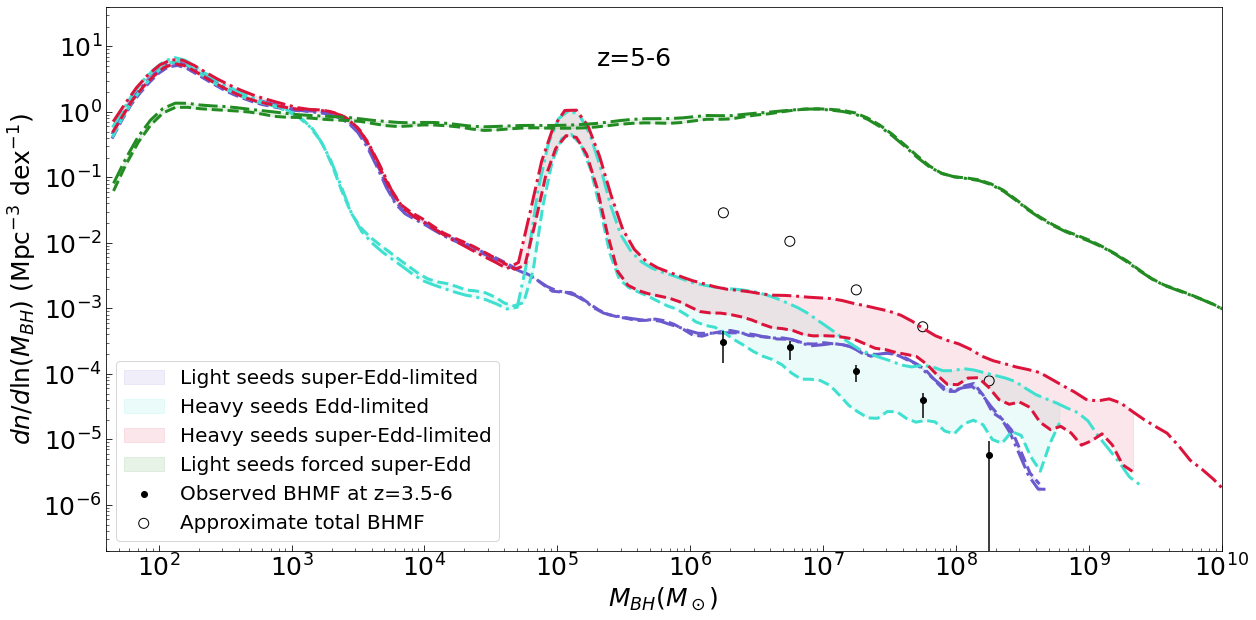}
    \includegraphics[width=0.9\textwidth]{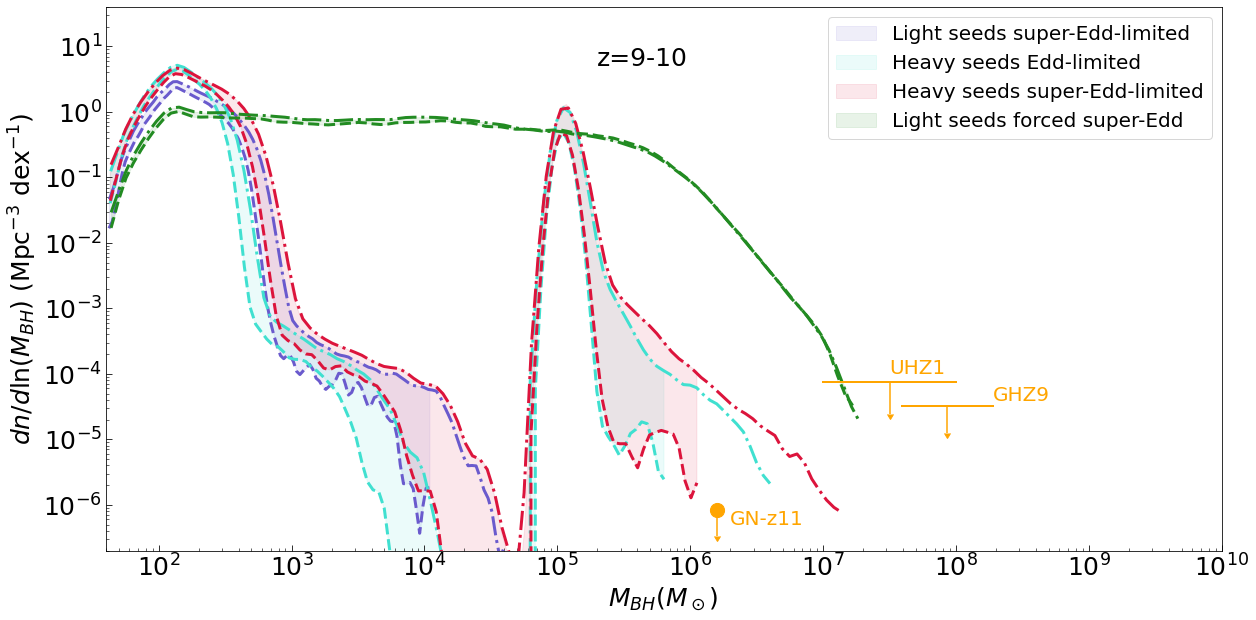}
    \caption{BHMF at $z=5-6$ ({\it top}) and $z=9-10$ ({\it bottom}), as predicted by our models. The default model with $v_{\rm BC}=0.8$ is shown as dashed lines and the ensemble model from the weighted mean of runs with $v_{\rm BC}=0, 0.8,2$ is shown with dot-dashed lines. The change in the streaming velocity $v_{\rm BC}$ only significantly affects heavy seed formation. We also show a model where the critical LW flux for DCBH formation is increased to $J_{\rm crit}=3,000$, to demonstrate that the peak amplitude for the heavy seed models are strongly dependent on model parameters. We compare our models against the observed BLAGN BHMF at $z=3.5-6$ \citep{Taylor2024,Matthee2023} and approximate upper limits derived from individual AGN observations at $z\sim10$ \citep{Maiolino2023,Bogdan2023,Fujimoto2023,Napolitano2024,Treu2022}. As the observed BHMF is specifically targeting broad-line AGN, we also plot (open circles) the approximate total BHMF, which includes obscured and dormant SMBHs, based on the TRINITY model \citep{Zhang2023}. All models with Bondi accretion can approximately reproduce the observed BHMF. As the accretion duty cycle is a free parameter here, and can be set to a lower value, the super-Eddington limited models represent the optimistic upper limit for the BHMF. However, only the forced super-Eddington model can reproduce the extreme UHZ1 system at $z=10.1$, observed at a redshift comparable to the range modeled here. Therefore, while the overall SMBH/AGN population can be produced under multiple scenarios, select special cases at high redshifts ($z\sim10$) can only emerge under the most extreme conditions that allow for very efficient accretion and/or heavy seed formation. Only a subset of SMBHs could have accreted so efficiently, however, as otherwise many more luminous AGN would have been observed at high redshifts beyond the current census.}
    \label{fig:bhmf}
\end{figure*}

We first examine the overall demographics of the high-redshift BH population, arising from each model, and show how model parameters can be calibrated. In Figure~\ref{fig:bhmf}, we show the resulting BHMF for all models at $z=9-10$ and $z=5-6$. We do not extend our models to $z\lesssim5$ as A-SLOTH was calibrated against observational constraints for a cosmologically representative galaxy population only at high redshifts $(z>4.5)$ \citep{Hartwig2022,Hartwig2024}.\footnote{Low-$z$ constraints for Milky Way-like galaxies are also considered in the calibration of A-SLOTH. However, it is unknown if the calibration results can be applied to cosmological simulations at $z < 4.5$, since they may be biased by the specific assembly histories of Milky Way-like galaxies.} The BHMF exhibits peaks at the BH seeding masses, at $\sim100$ M$_\odot$ for light seeds, and $10^5$ M$_\odot$ for heavy seeds, respectively. The amplitude of the (heavy-seed) peak is very sensitive to model parameters (see in particular Section~2.2 on the dependence on $J_{\rm crit}$), and we here probe the upper limit on heavy-seed (DCBH) abundance with our assumption for the strength of local LW sources. We specifically consider a model where the critical LW value for DCBH formation is increased to $J_{\rm crit}=3,000$ (with $v_{\rm BC}=0.8$), which results in a much lower peak amplitude, thus demonstrating the strong dependence on heavy-seed criteria. We note that the amplitude of the fiducial DCBH peak ($\sim1-0.1$ Mpc$^{-3}$ dex$^{-1}$) is higher than locally inferred abundances for $10^6$\,M$_\odot$ SMBHs ($\sim10^{-2}-10^{-3}$ Mpc$^{-3}$ dex$^{-1}$) \citep{Greene2020,Marconi2004}. Such a large DCBH population could possibly be accounted for if many heavy seeds remained dormant and were not luminous, either through effects of stellar feedback \citep{Jeon2023} or dynamical wandering away from galaxy centers \citep{Mezcua2020,Reines2020}, thus rendering them undetectable with current observational facilities. Furthermore, our heavy-seeding model represents upper limits, in line with the conclusion in other works, invoking similarly optimistic scenarios, that heavy seeds could account for all SMBHs in the Universe under these conditions \citep{Chon2024}. Lastly, the heavy seed model had originally been introduced to explain the most massive SMBHs at early times, similar to the argument presented here that they are needed to account for the extremely massive outlier cases (see below). However, they are in general not necessary to reproduce the $z=3.5-5$ BHMF, such that the conditions for heavy seed formation could be more restrictive to reduce their formation and peak height without affecting our overall conclusions.

At $z=9-10$, we compare our models against select AGN observations at comparable redshifts, UHZ1 at $z=10.1$ \citep{Bogdan2023}, GHZ9 at $z=10.145$ \citep{Napolitano2024}, and GN-z11 at $z=10.6$ \citep{Maiolino2023}. As can be seen, under our parameter choices, all Bondi accretion models roughly match the $z=3.5-6$ BLAGN BHMF in slope and amplitude, except that the super-Eddington limited models are slightly above the observations at the highest BH masses. Such over-prediction can be mitigated by decreasing the accretion duty cycle, which may be lower for highly accreting objects. Thus, our super-Eddington limited models represent somewhat optimistic, but still physically plausible estimates for the BHMF.

Conversely, at $z\sim10$, only the forced super-Eddington accretion model for light seeds is able to match the upper limit on volume density inferred from UHZ1 \citep{Fujimoto2023}. We note that even super-Eddington-limited heavy seeds could not reproduce the inferred UHZ1 abundance at $z\sim10$. However, the forced super-Eddington model would clearly overproduce the BHMF, compared with the $z=3.5-6$ observations \citep{Taylor2024,Matthee2023,Kocevski2023}. This model reaches the physical upper limit of accretion, where the BH has accreted all/most of the cold gas available in the halo. In contrast, the volume density inferred for the less-massive GN-z11 SMBH can be achieved with heavy seeds. Furthermore, without forced super-Eddington accretion, all light seeds remain below $10^5$ M$_\odot$ throughout, and thus cannot produce the observed SMBHs at $z\sim10$. 

\subsection{Black hole mass evolution}

\begin{figure}[htb!]
 \includegraphics[width=0.5\textwidth]{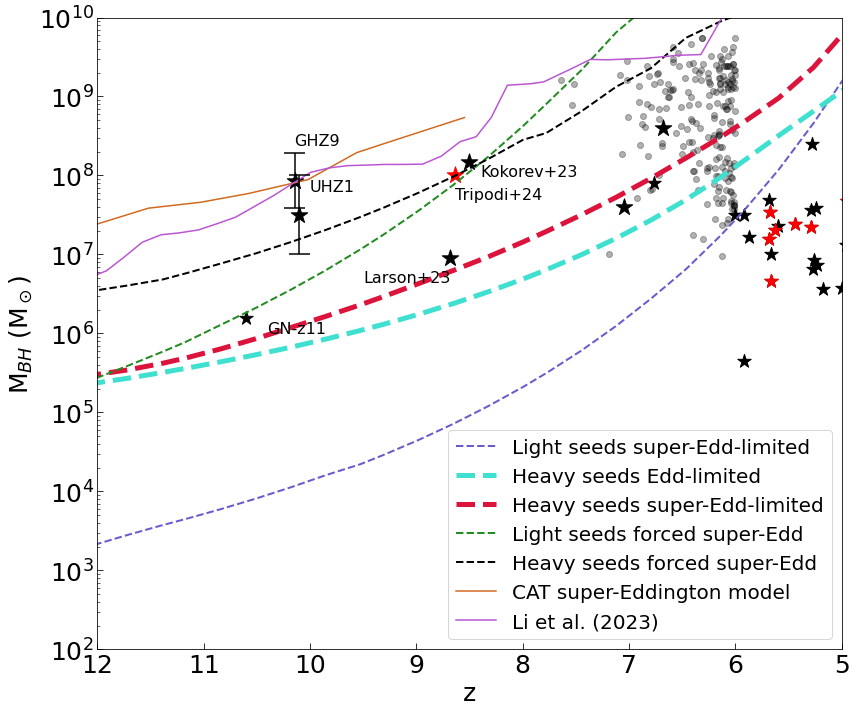}
  \caption{Most massive SMBH vs. redshift for different seeding/growth models with $v_{\rm BC}=0.8$. We show estimated BH masses for select LRDs (red stars), \citep{Taylor2024,Tripodi2024}, adopting the AGN interpretation. We further plot select \textit{JWST}-observed high-redshift AGN as black stars \citep{Larson2023,Bogdan2023,Maiolino2023_2,Kokorev2023,Furtak2023,Juodbalis2024,Napolitano2024}, and high-redshift quasars as black dots \citep{Wang2010,Willott2017,Decarli2018,Izumi2018,Pensabene2020,Inayoshi2020,Fujimoto2022}. We compare our model against the CAT SAM heavy seed super-Eddington model \citep{Schneider2023}, and a heavy seed model with growth constrained through the quasar luminosity function \citep{Li2023}. The highest-redshift ($z\gtrsim8$) detections can only be reproduced with the forced Eddington accretion model, whereas the lower-redshift AGN ($z\sim6$) can also be produced by the Bondi-accretion, Eddington-limited models. To be able to produce the massive UHZ1 object, we include a model with heavy seeds and the forced Eddington growth model, albeit at a lower duty fraction of 0.5. Therefore, the most massive AGN at the highest redshifts may have formed under the extreme conditions that allowed for continuous Eddington/super-Eddington accretion, while lower redshift AGN could have formed under more common, less extreme conditions.}
    \label{fig:bhmost}
\end{figure}

For each of the models, we further examine the mass of the most massive BH at each redshift and compare it against existing high-redshift quasar and \textit{JWST} AGN observations. In Figure~\ref{fig:bhmost}, we show such a comparison. Most models fail to reproduce the most-massive AGN observed by {\it JWST} at $z\gtrsim8$, in particular the peculiar UHZ1 and GHZ9 systems, as well as the most massive, high-$z$ quasars, with the exception of the forced-Eddington cases. The latter in turn are close to the ``causal limit'', where all/most of the cold gas supply in the host halo is accreted. We note that to account for UHZ1, such sustained super-Eddington accretion onto heavy seeds would be required. The extreme accretion and resulting growth of the forced-Eddington models will need to peter out at later times, because if these growth trajectories were to continue, the resulting SMBHs would acquire masses of $\sim10^{10}$\,M$_\odot$ by $z\sim6$, while such massive objects are extremely rare \citep{Wu2015}. This could be explained by the denser environments at higher redshifts that are more suited to enable extreme accretion, whereas gaseous conditions at lower redshifts become less dense and more readily affected by stellar and BH feedback, resulting in less efficient SMBH accretion \citep[e.g.,][]{Feng2014,Ni2022}.

Our findings agree with previous theoretical predictions \citep{Inayoshi2020,Volonteri2010} and SMBH observations prior to \textit{JWST} \citep{Banados2018,Yang2021}, concluding that to explain the most massive SMBHs in the early Universe, non-standard pathways such as heavy seeds and/or super-Eddington accretion are required. Our results are further consistent with other theoretical models, such as those obtained within the Cosmic Archaeology Tool (CAT) SAM \citep{Schneider2023}, where a heavy seed together with merger-driven super-Eddington accretion was needed to match UHZ1, or the model where a heavy seed with growth constrained through the quasar luminosity function is invoked \citep{Li2023}. Overall, the ``most-massive BH diagnostic'' is quite constraining for early seeding and growth models, and the prospect of extending this frontier to even higher redshifts with upcoming, ultra-deep {\it JWST} observations is compelling. Thus extending the high-redshift frontier may be effectively assisted by gravitational-lensing magnifications, such as with the ongoing GLIMPSE survey \citep{Kokorev_GLIMPSE_2024}.

\begin{figure*}
\gridline{
\fig{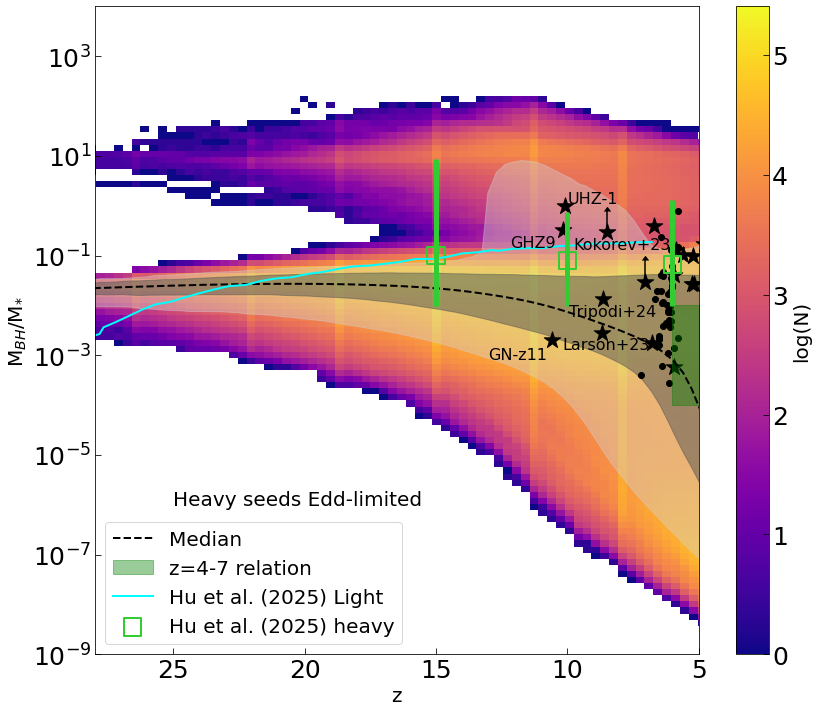}{0.5\textwidth}{}
\fig{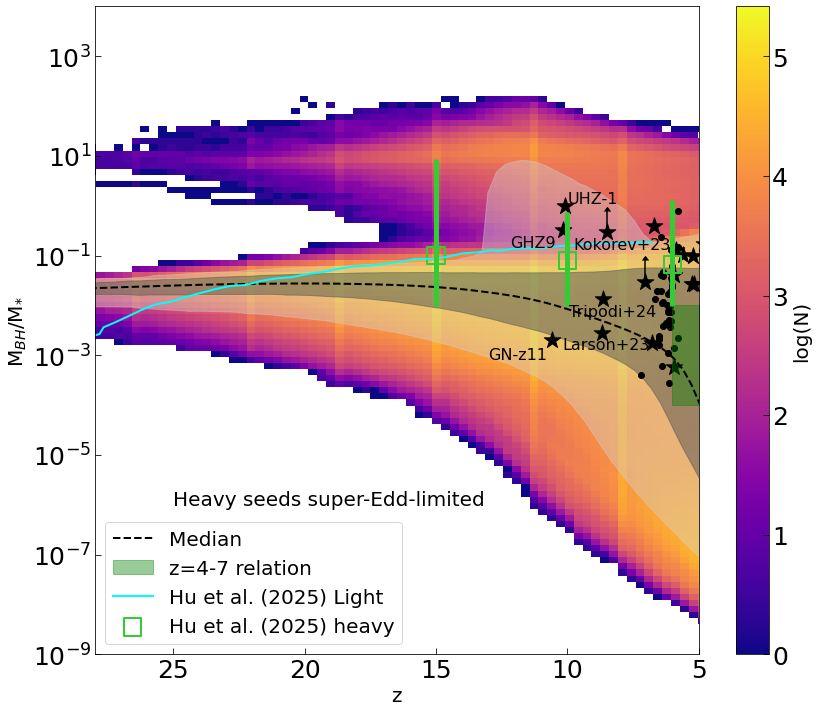}{0.5\textwidth}{}
}
\gridline{
\fig{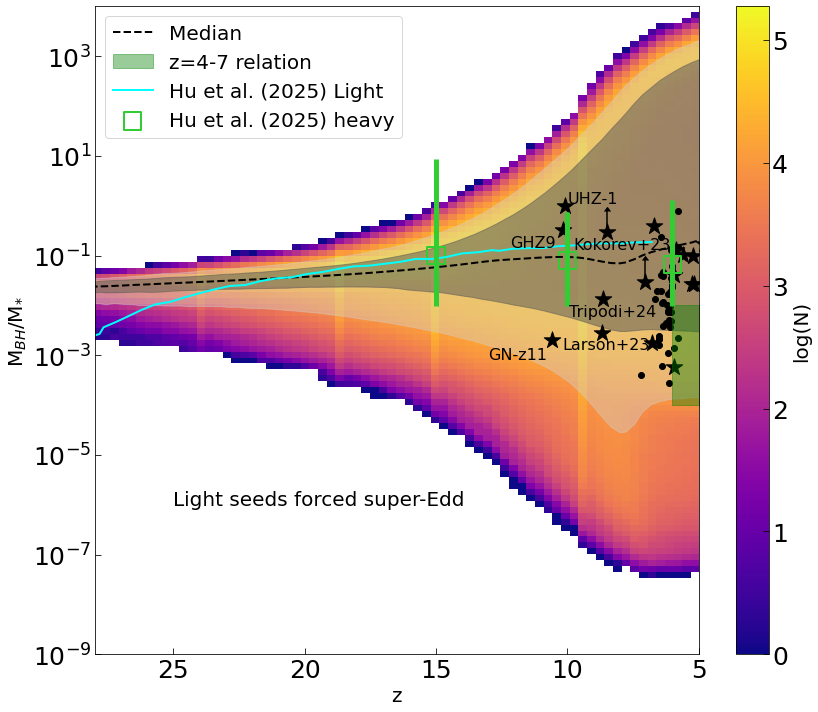}{0.5\textwidth}{}
\fig{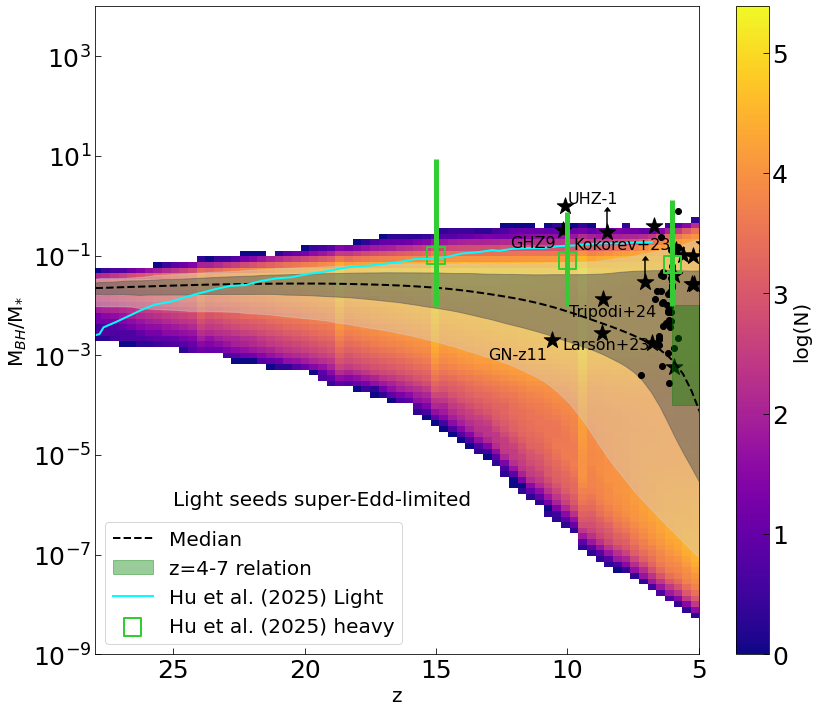}{0.5\textwidth}{}
}
    \caption{Co-evolution of SMBH and its host system. We show the ratio of SMBH to stellar mass vs. redshift, indicating the prevalence of cases (number $N$) with the given color-coding convention. For each model, we also indicate the median ratio with the black dashed line and the weighted 1- and 2-$\sigma$ spreads with the ochre green and transparent white shaded regions. Specifically, all models assume a `standard' value for baryon-DM streaming of $v_{\rm BC}=0.8$. We further plot select values for \textit{JWST}-observed, high-redshift AGN as black stars \citep{Larson2023,Bogdan2023,Maiolino2023_2,Kokorev2023,Furtak2023,Juodbalis2024,Tripodi2024,Napolitano2024}, and high-redshift quasars as black dots \citep{Wang2010,Willott2017,Decarli2018,Izumi2018,Pensabene2020,Inayoshi2020}. For additional context, we reproduce the BH to stellar mass relation inferred at $z\sim4-7$ \citep{Pacucci2023}, with the shaded (green) region at the right-end of the panels. We compare our results against select heavy and light seed models of \citet{Hu2025}, in which heavy seeds form later. All models produce a wide range of BH to stellar mass ratios at $z\lesssim10$, covering most of the observations. Thus, it will be difficult to distinguish between BH evolutionary pathways, based on the BH to stellar mass observations at $z \lesssim10$ alone. However, clear differences arise at even higher redshifts ($z\gtrsim15$), to be probed with future ultra-deep {\it JWST} observations.}
    \label{fig:bhratio}
\end{figure*}

\subsection{Overmassive black holes}

Regarding the co-evolution of SMBHs and their host systems, {\it JWST} has established the key result that many of the newly discovered sources at high-$z$ are overmassive, where the SMBH to galaxy stellar mass ratio is much higher than in the local Universe \citep[e.g.,][]{Maiolino2023_2}. To explore this complex co-evolution within our semi-analytical modeling, we show the ratio of BH to stellar mass for the different cases across redshifts in Figure~\ref{fig:bhratio}. As is evident in the figure, most models can explain the range of observed ratios at $z\lesssim 10$ including the extreme UHZ1 case. This result agrees with previous theoretical studies, including the CAT SAM \citep{Trinca2024}, as well as other theoretical models that do not explicitly invoke the heavy seed formation channel \citep{Hu2025}. In fact, while heavy seeds are considered in \citet{Hu2025}, the BH-to-stellar mass ratios for light and heavy seeds become indistinguishable at the observed redshifts. Similarly, according to the BRAHMA simulations \citep{Bhowmick2024}, the details of the heavy seeding conditions are found to not have a large impact on the resulting distribution of mass ratios. Given the optimistic assumptions in our heavy seeding model, in terms of their formation and growth, our results should be considered as upper limits for the heavy seed pathway. When considering all seeding channels, heavy seeds represent a small fraction of the overall SMBH population, and only select cases are extremely overmassive as seen in the trend of the median BH to stellar mass ratio in Fig.~\ref{fig:bhratio}. Even for models with heavy seeds, the median ratio evolves to lower values at later times. This is in line with existing observations, where few systems are highly overmassive, but many high-redshift AGNs exhibit lower $M_{\rm BH}/M*$ ratios. The notable exception is the light-seed model with (super-Eddington-limited) Bondi accretion (bottom-right panel), where typical ratios of $M_{\rm BH}/M_{\ast}\sim 10^{-2}$ are established, in line with earlier results that light-seed models cannot efficiently grow, unless sustained periods of super-Eddington accretion can occur \citep[e.g.,][]{Jeon2023}. We further note that the `Light seeds forced super-Edd' model produces extremely overmassive systems $(M_{\rm{BH}}/M_*\sim10^3)$ in large number at $z\sim5$. However, this model, as stated earlier, is an extreme case and reaches the physical upper limit of accretion, where all available cold gas is accreted. Therefore, such a model where all seeds grow in this hyper-efficient way until lower redshifts $(z\sim5)$ is not plausible. The BHMF is overproduced in this model at lower redshifts as well (Fig.~\ref{fig:bhmf}), further confirming this conclusion. We discuss below that to explain current observations, only a small subset ($0.01\%$ at $z\sim5-6$) of cases could follow such extreme growth trajectories, especially at lower redshifts (See Section \ref{sec:discussion}).

Intriguingly, the near-degeneracy of light- and heavy-seed models can be broken when pushing to even higher redshifts. As can be seen, at $z\gtrsim 15$ the heavy-seed models exhibit a bifurcation into two separate branches, overmassive and ``normal'', at ratios of $\sim 10$ and $10^{-2}$, respectively. The light-seed growth pathways with forced super-Eddington accretion, on the other hand, initially show a narrow range of BH to stellar mass ratio, and only later on, at $z\lesssim 15$ extend into the overmassive domain. If this qualitatively different behavior at high redshifts can be probed with ultra-deep {\it JWST} surveys, we may have a tell-tale signature of light vs. heavy SMBH seeding. Next, we will further discuss ways to address this key challenge of distinguishing between early seeding and growth channels.

\section{Signature of BH Seeding Pathways} \label{sec:discussion}

From the above results, we conclude that while the observed AGN population and select massive objects at $z\lesssim8$ can be fairly well described by all our models, there are differences in the most extreme objects the models can produce, as well as the overall amplitude of the BHMF, and the location of specific BHMF peaks. We note that larger differences arise at higher redshifts $(z\gtrsim9)$, providing a greater potential to empirically distinguish between models. Specifically, when heavy seeds exist, there is a peak in the mass function at the mass where heavy seeds are initially formed at ($10^5$ M$_\odot$, for our assumption here), as seen in Fig.~\ref{fig:bhmf}.

\begin{figure*}[!htb]
    \centering
    \includegraphics[width=0.9\textwidth]{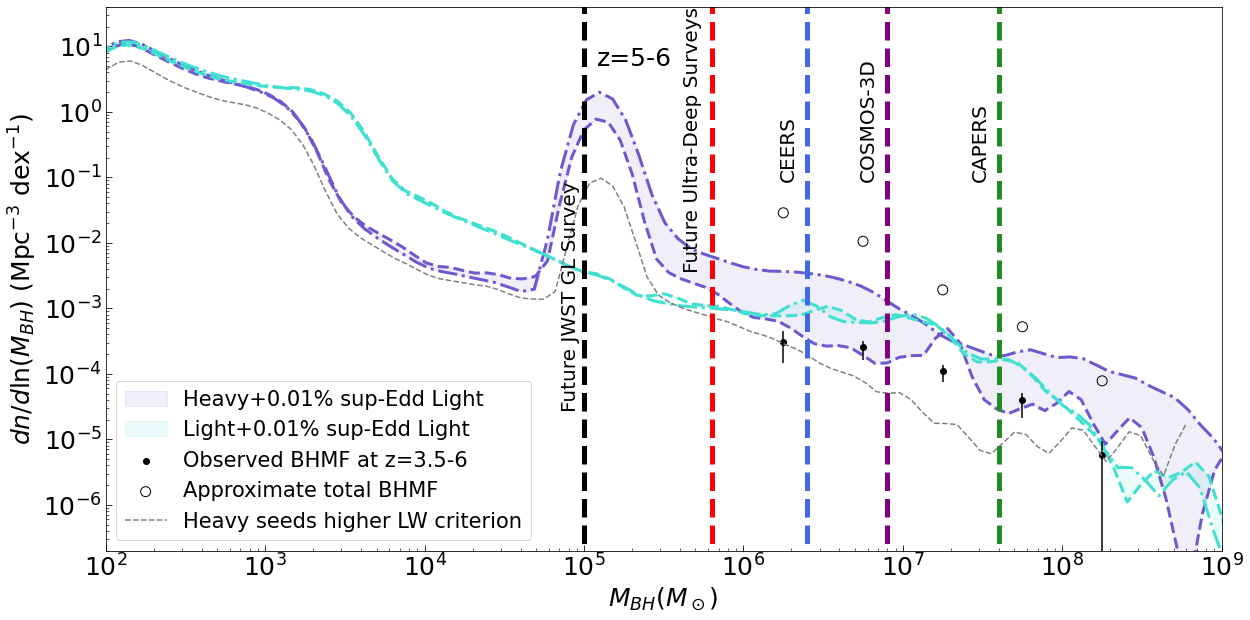}
    \includegraphics[width=0.9\textwidth]{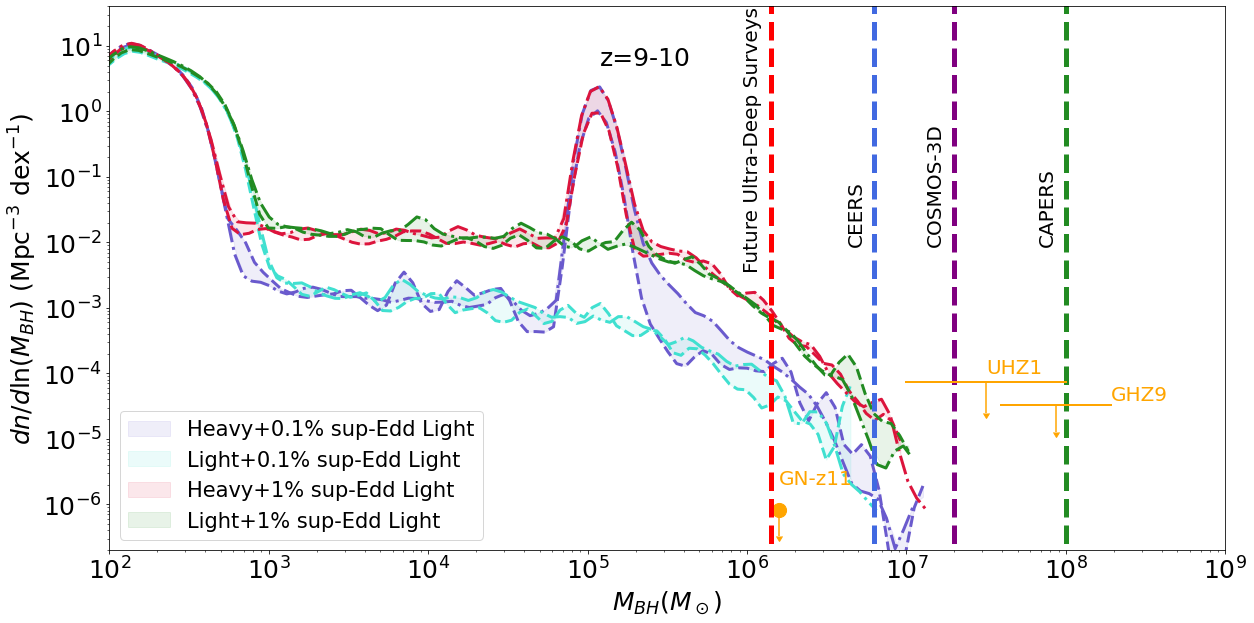}
    \caption{BHMF of various model combinations at $z=5-6$ and $z=9-10$, similar to Fig.~\ref{fig:bhmf}. Specifically, we combine the `Heavy seeds Edd-limited' or `Light seeds super-Edd-limited' models with a small fraction of `Light seeds forced super-Edd' model by randomly choosing BHs from each model, with a small probability (0.01-1\%) of choosing the forced super-Edd BHs. At $z=5-6$, the super-Edd fraction is 0.01\% to match the observed BHMF, while at $z=9-10$, we vary the fraction between 0.1 and 1\%. The dashed line is the model with $v_{\rm BC}=0.8$, and the dot-dashed one the model with the weighted mean of $v_{\rm BC}$ values. The combination of a few BH seeds accreting very efficiently with a majority of seeds accreting at less extreme rates can reproduce all high-redshift observations so far. When heavy seeds are included, a distinctive peak near the DCBH seed mass arises ($10^5$\,M$_\odot$ for our models), persisting across redshifts ($z\lesssim 15$). JWST gravitational lensing surveys, such as spectroscopic follow-up to the GLIMPSE survey \citep{Atek2023}, could detect such a peak at $z\sim6$. However, without lensing, current and future \textit{JWST} surveys such as CEERS \citep{Finkelstein2025}, COSMOS-3D \citep{Kakiichi2024}, and CAPERS \citep{Dickinson2024}, will likely not be able to directly detect this DCBH peak, as indicated by their BH mass detection limits, shown with the vertical dashed lines (See Section~\ref{sec:discussion}). However, future ultra-deep surveys that can probe up to $M_{\rm BH}\sim10^6$\,M$_\odot$ at $z\sim9$ will be able to determine the fraction of highly-efficient accreting BHs at higher number densities, up to two orders of magnitude than current surveys, resulting in smaller uncertainties. The volume density of the massive AGN sensitively depends on the fraction of high-accreting BHs, such that the number of host halos/galaxies that are able to support such extreme growth could be constrained with near-future observations. The prevalence of heavy seeds could be indirectly constrained as well, if future surveys show a much lower number density than predicted when super-Eddington seeds are included (closer to the limit set by GN-z11), such that the lower number densities predicted for the Eddington-limited heavy seed scenario would provide a better fit (see also Fig.~\ref{fig:bhmf}).}
    \label{fig:bhmf_combine}
\end{figure*} % have a range of percentage to show?
% luminosity function?

Furthermore, under the standard Bondi accretion, the massive SMBHs at $z\sim10$ with masses larger than $10^5$ M$_\odot$ all originate from heavy seeds with no light seeds able to grow that massive at early times. Long periods of efficient super-Eddington accretion, as in our ``Light seeds forced super-Edd'' model, are necessary for light seeds to produce such massive cases. We thus confirm that extreme growth may be necessary to produce objects similar to UHZ1. 

However, determining which BH seeds could accrete at super-Eddington rates is difficult. Small-scale high-resolution simulations have demonstrated that such extreme conditions are possible \citep[e.g.][]{Gordon2025,Jiang2019,Jiang2019_2,Kaaz2024,Hu2022}, but in larger-box simulations with lower resolution, such conditions cannot be easily identified. Even more so in this work, we cannot determine which halos experience super-Eddington conditions, since we do not explicitly model the detailed halo gas structure within our semi-analytical framework. Therefore, we instead approximately assume that some fraction of all BH seeds will accrete at extreme rates. In Figure~\ref{fig:bhmf_combine}, we show the BHMF arising from the combination of two modes: forced super-Edd and (super-)Edd-limited. 
We attempt to mimic the real Universe with these hybrid models, where most BHs accrete inefficiently represented by the Heavy seeds Edd-limited model or the Light seeds super-Edd-limited model, while a few exist in environments that allow very efficient accretion as in the case of Light seeds forced super-Edd. At higher redshifts where denser environments are more common, such highly accreting BHs are assumed to be more common as well. Thus, the forced super-Eddington model, assumed to contribute 0.01\% of BHs, can reproduce the observed BLAGN BHMF at $z=5-6$, in combination with and without heavy seeds. At $z=9-10$, with fewer constraints on the BH population, we consider two contribution values from the efficiently accreting BHs, 0.1 and 1\%. 

When heavy seeds are included in this combination of models, both the BLAGN BHMF at $z\sim5$ and the UHZ1 constraint can be reproduced, considering that the latter is an upper limit, and that the actual number density of UHZ1-class systems may be significantly lower. With just light seeds, the earliest AGN systems can only be reproduced when including a fraction of super-Eddington accretion. As before in Fig.~\ref{fig:bhmf}, the key difference when heavy seeds exist is the prominent peak at the heavy seeding mass. Beyond that, the predicted BHMF slope is very similar at $M_{\rm BH}\gtrsim 10^6$\,M$_{\odot}$ between the two scenarios. The limiting AGN mass allowing detection with current and future \textit{JWST} surveys \citep{Finkelstein2025,Kakiichi2024,Dickinson2024} at $z\sim9-10$ robustly extends only down to $\sim10^6$ M$_\odot$. Therefore, based on the BHMF alone, we conclude that future {\it JWST} surveys may still not be able to directly distinguish between seeding pathways for the first SMBHs. We compute these estimated AGN black hole mass detection limits by simulating \textit{JWST} observations of broad-line AGN using the \texttt{Pandeia} \textit{JWST} ETC engine \citep{pontoppidan16}. We use the relations given in \cite{Reines2013} to model broad H$\alpha$ (or broad H$\beta$ at $z>7$) lines emitted by BLAGN of varying black hole mass. We then use \texttt{Pandeia} to simulate how these idealized model lines will appear when observed by different \textit{JWST} programs, and determine---for each program, as a function of redshift---the limiting black hole mass at which the broad line can no longer be robustly recovered when fit with Bayesian techniques. We here do not consider the limiting effect of a survey's effective area coverage in observing the most massive AGNs, as in the context of this work, we aim to constrain whether these surveys will be able to detect more common and thus less massive SMBHs.

There are, however, promising strategies to work around this conclusion. Gravitational lensing could provide the flux magnification to enable the discovery of the less massive SMBHs at $\sim10^5$ M$_\odot$, which are otherwise too faint to be observed \citep{Jeon2023}. While gravitational lensing is the only way to probe these low-mass black holes with reasonable integration time on current facilities, the survey volumes of lensed fields are significantly smaller. This effect could be partially offset by the increased number density of faint black holes, but heroic integrations are required to reach these faint sources. One example is the recently approved Director's Discretionary Time (DDT) follow-up observations of the Abell~S1063 GLIMPSE field \citep{Kokorev_GLIMPSE_2024,Fujimoto2025}. With 40 hour integrations and medium spectral resolution, these upcoming observations will be sensitive to black hole masses down to $10^5~M_\odot$, sufficient to provide the first empirical probe of the peak of the BLAGN BHMF due to the DCBH seed mass, although the height of the peak is not fully constrained, subject to the DCBH formation criteria (see Section~\ref{sec:demographics}). 

Alternatively, the fraction/number of highly accreting SMBHs could be well constrained through \textit{JWST} observations in the near future. The existing $z\sim10$ AGN observations are most likely the extremely bright objects in that period. When ultra-deep surveys are able to find more such objects at $z\sim10$, their number density can be constrained with improved accuracy. Moreover, even without gravitational lensing, future ultra-deep \textit{JWST} spectroscopic observations of blank fields could further detect SMBHs at $M_{\rm BH}\sim10^6$ M$_\odot$ via broad H$\beta$ or H$\gamma$ observations. As shown from our models in Fig.~\ref{fig:bhmf_combine}, the fraction of highly accreting SMBHs affect the amplitude of the BHMF. If the BHMF at $z\sim10$ can be constrained at lower BH masses, the fraction of halos/galaxies that can support such highly accreting objects can also be inferred.

It can be argued that the duty fraction we use for the forced super-Eddington model, 0.8, is an extreme value, especially as super-Eddington SMBHs are expected to have a low duty fraction \citep{Fontanot2023,Pezzulli2017,Trinca2024}. However, degeneracy exists between the super-Eddington accretion rate and the accretion duty fraction. The forced super-Eddignton rate could be set to a higher value to increase the overall BHMF as well. We have tested decreasing the duty fraction (0.1) but increasing the forced super-Eddington rate (12). The resulting BHMFs were nearly identical to the forced super-Eddington model with no significant differences. This is expected, as in principle, both the duty fraction and the super-Eddington rate work to moderate the accretion rate in our model (see Eq.~\ref{eq:macc} and \ref{edd_acc}). Thus, our forced super-Eddington model represents the scenario of extreme BH growth, either through high duty fraction or extreme super-Eddington accretion. Furthermore, for the forced super-Eddington case, many BHs in the model accrete all available cold gas in the halo, reaching the physical upper limit of accretion\footnote{We have estimated in post-processing the Bondi-Hoyle boost factor, $\alpha$, that would be needed to reproduce the forced-Eddington accretion rates. For the top 25\% most-massive SMBHs, equivalent boost factors would be $\sim10^3-10^4$. Thus, the forced-Eddington model probes the extreme upper limit of BH growth.}. This limit of available gas is a robust limit even when the detailed physical process of BH accretion is not known. Therefore, the super-Eddington fraction we consider here is close to the most optimal SMBH growth, and so our models do show the differences that will exist in the BHMF depending on the fraction of the extreme efficiently accreting SMBHs.

The CEERS survey has already reached the massive end of the BHMF, and can thus begin to measure the highly-accreting SMBH fraction. However, CEERS is now completed, yet only a handful of AGN detections at $z\sim10$ exist, insufficient to constrain this fraction well. As the most massive and brightest SMBHs will be the rarest and lowest in volume density, such a result is not surprising. However, future \textit{JWST} ultra-deep surveys will be able to probe less massive and more abundant SMBHs, up to $\sim$2 orders of magnitude higher in volume density according to our models. Numerous additional AGN are thus expected to be discovered in the near-future, and the fraction of efficiently accreting SMBHs will be constrained much more strongly.

Finally, the existence of heavy seeds could be indirectly demonstrated through future surveys. All models summarized in Figure~\ref{fig:bhmf_combine} predict values above the upper limit set by the volume density of GN-z11 at $z\sim10$. This could be due to observational incompleteness, missing fainter or obscured AGN. However, if future observations were to show that the SMBH BHMF is closer to the GN-z11 value, this would indicate that very few SMBH would experience efficient growth at (sustained) super-Eddington levels. Similar conclusions could be drawn if the current high upper limit for the abundance of UHZ1-type systems would be revised downwards. Since light seeds without extended super-Eddington growth cannot reach masses as high as the GN-z11 SMBH at $z\sim10$, and if observations were to confirm the low abundance estimate in this mass range, a heavy seed origin would be favored. UHZ1 and GHZ9 in this case would be extreme outliers, which may require heavy DCBH seeds, possibly combined with efficient super-Eddington growth episodes. Other theoretical models like TRINITY, constrained with high-redshift UV luminosity functions, could not reproduce a system like UHZ1 and have also found it to be an outlier case \citep{Zhang2023}. The constraints on such extreme and massive SMBHs will be rendered much stronger with the Roman telescope, which is expected to find massive quasars at $z\sim6-10$ \citep{Zhang2024}.

% discern with luminosity function?

\section{Summary and Conclusions} \label{sec:conclusions}

In this work, we have modified the semi-analytic model A-SLOTH to include the seeding and evolution of the first SMBHs in the Universe. We explore various seeding and accretion scenarios, including heavy DCBH seeds, light stellar remnant seeds, Bondi-Hoyle accretion, and enforced super-Eddington accretion. We find that, even with differences in the BHMF features and overall amplitude, the observed BLAGN BHMF at $z=3.5-6$ can be largely reproduced with a broad selection of models and their parameters, albeit with considerable degeneracies between them, similar to constraints from more local BHMF determinations \citep{Evans2025}. 

To possibly break this degeneracy, we examine our models at higher redshifts, $z\sim9-10$, and find that although existing and near-future \textit{JWST} surveys may still not be able to directly identify the dominant SMBH seeding scenario, powerful empirical constraints can be obtained. Both seeding models are able to produce the massive AGN and overmassive systems observed at high redshifts, either through having higher mass initially (heavy seeds) or accreting efficiently at super-Eddington rates (light seeds). A key target for the next cycles of {\it JWST} observations is to constrain the SMBH accretion mode, as they will preferentially discover the extremely high-accreting AGN. The amplitude of the BHMF will change according to the number of efficiently accreting AGN, which are the main sources for the extreme objects currently being observed at $z\sim10$, whether they originate from heavy or light seeds. Future ultra-deep surveys with \textit{JWST} will be able to observe SMBHs with masses as low as $\sim10^6$\,M$_\odot$, which are predicted to be around two orders of magnitude more abundant than currently observable SMBHs so that their super-Eddington fraction will be much better constrained. 

Furthermore, the existence of heavy DCBH seeds could be indirectly confirmed through observations of SMBH to stellar mass ratios at $z\gtrsim15$ or if the observed BHMF at mass ranges around $10^6$ M$_\odot$ in future surveys exhibits lower abundances than predicted when a super-Eddington accretion mode is included. Put differently, any prevalence of (sustained) super-Eddington accretion modes would drive up the number densities of massive SMBHs, because such efficient accretion would boost a significant fraction of the abundant light seeds into the observable regime. Conversely, without super-Eddington accretion, only heavy seeds can reach high enough masses to be observable by $z\sim10$, so that a low BHMF-amplitude observation will imply that the observed SMBHs originated from heavy seeds. The existence of heavy seeds could even be directly confirmed through \textit{JWST} gravitational lensing surveys. In addition, different pathways for heavy seed formation could leave different observational signatures \citep{Bhowmick2025}, as we have only tested one heavy seed formation mechanism in this work.

Finally, complementary future multi-wavelength and multi-messenger observations could further elucidate SMBH seeding pathways. Pulsar timing array observations have detected the stochastic stochastic gravitational wave background \citep{Agazie2023,Antoniadis2023,Reardon2023,Xu2023}, which if sourced from binary SMBHs \citep{Hobbs2017,Romano2017}, could be used to constrain their population at high redshifts. % \citep[e.g.,][]{Liu2024}. 
Future GW observatories like LISA will be able to more robustly detect such signals \citep{Robson2019}. Furthermore, future X-ray missions could detect additional AGN at earlier times, targeting sources which may be too X-ray weak to be detected currently \citep{Kocevski2024,Yue2023,Juodbalis2023}. E.g., the Athena mission will be able to detect much fainter AGN than currently possible with existing optical and near-IR surveys \citep{Barret2013}, and the AXIS mission will discover SMBHs with masses below $10^5$ M$_\odot$ to probe SMBH seeding pathways \citep{Reynolds2023,Cappelluti2024}. In the long-term future, the Lynx X-ray mission aims to detect the first BH seeds with a 100 times increase in X-ray sensitivity compared to the Chandra observatory \citep{Gaskin2019}. Therefore, with \textit{JWST} observations charting the broad outlines of the first SMBHs and their evolutionary pathways, future observatories will be able to follow up in greater depth, thus completing our understanding of how the Universe has created these massive objects so early in its history.


\begin{acknowledgments}
The authors acknowledge the Texas Advanced Computing Center (TACC) for providing HPC resources under FRONTERA allocation AST22003. BL gratefully acknowledges the funding of the Royal Society University Research Fellowship and the Deutsche Forschungsgemeinschaft (DFG, German Research Foundation) under Germany's Excellence Strategy EXC 2181/1 - 390900948 (the Heidelberg STRUCTURES Excellence Cluster). AJT acknowledges support from the UT Austin College of Natural Sciences.
\end{acknowledgments}

%\vspace{5mm}


%\software{}




\bibliography{ms}{}
\bibliographystyle{aasjournal}



\end{document}